\documentstyle[12pt,graphicx]{article}
\begin{document}
\title{Nonrelativistic Charged Particle-Magnetic Monopole Scattering in the 
Global Monopole Background}
\author{A. L. Cavalcanti de Oliveira \thanks{E-mail: alo@fisica.ufpb.br} \ 
and E. R. Bezerra de Mello \thanks{E-mail: emello@fisica.ufpb.br}
\\
Departamento de F\'{\i}sica-CCEN\\
Universidade Federal da Para\'{\i}ba\\
58.059-970, J. Pessoa, PB\\
C. Postal 5.008\\
Brazil}
\maketitle
\begin{abstract}
We analyze the nonrelativistic quantum scattering problem of a charged particle by an Abelian magnetic monopole in the background of a global monopole. In addition to the magnetic and geometric effects, we consider the influence of the electrostatic self-interaction on the charged particle. Moreover, for the specific case where the electrostatic self-interaction becomes attractive, charged particle-monopole bound system 
can be formed and the respective energy spectrum is hydrogen-like one.

\end{abstract}

\newpage

\section{Introduction}

In this paper we analyse the scattering problem of a massive charged particle by an Abelian magnetic monopole in the background of a global monopole. In order to develop this analysis we shall consider that both defects are at the same position.

The most elegant formalism to describe the Abelian pointlike magnetic monopole has been developed by Wu and Yang\cite{Wu-Yang}. They defined the vector potential $A_\mu$ in two overlapping regions, $R_a$ and $R_b$, which cover the whole space. Using spherical coordinate system, with the monopole at origin they have chosen:
\begin{eqnarray}
\label{Ra}
R_a&:& 0 \leq \theta < 1/2\pi+\delta, \  r>0, \  0\leq \phi 
<2\pi \ , \\
R_b&:& 1/2\pi-\delta< \theta \leq \pi, \ r>0, \ 0\leq \phi < 2\pi ,\\ 
\label{Rab}
R_{ab}&:& 1/2\pi-\delta< \theta< 1/2\pi+\delta, \ r>0,  \
0\leq \phi < 2\pi \ , 
\end{eqnarray}
with $0<\delta \leq \pi/2$, and $R_{ab}$ being the overlapping region.

The only non vanishing components of vector potential are
\begin{eqnarray}
\label{A}
(A_\phi)_a=g(1-\cos\theta) \ ; \ (A_\phi)_b=-g(1+\cos\theta) \ ,
\end{eqnarray}
$g$ being the monopole strength. In the overlapping region the non vanishing components are related by a gauge transformation
\begin{equation}
(A_\phi)_a=(A_\phi)_b+\frac ieS\partial_\phi S^{-1} \ ,
\end{equation}
where $S=e^{2iq\phi}$, $q=eg=\frac{n}{2}$ in units $\hbar=c=1.$

The global monopole is a heavy object formed in the phase transition of a system composed by a self-coupling scalar field triplet $\phi^a$ whose original global $O(3)$ gauge symmetry is spontaneously broken to $U(1)$ \cite{Barriola}.

The metric tensor corresponding to a pointlike global monopole spacetime is given by the line element below,
\begin{equation}
\label{GM}
ds^2=-d t^2+\frac{dr^2}{\alpha^2}+r^2(d\theta^2+\sin^2\theta d\phi^2) \ .
\end{equation}
The parameter $\alpha$ codifes the presence of the global monopole.

Because the global monopole spacetime presents a non trivial topology and curvature, a charged particle placed in this spacetime becomes subjected to an electrostatic self-interaction\cite{Mello}
\begin{equation}
U=K/r \ ,
\end{equation}
where $r$ is the distance from the particle to the monopole, $K=e^2S(\alpha)/2$ and $e$ is the charge of the particle. The numerical factor $S(\alpha)$ is a finite and positive number for $\alpha<1$ and negative for $\alpha>1$.

\section{Nonrelativistic Quantum Mechanical Analysis}

According to Wu and Yang\cite{Wu-Yang1}, the solution of the Schr\"odinger equation in such external field will not be an ordinary function but, instead, {\it section}, i.e., the solution assumes values $\Psi_a$ and $\Psi_b$ in $R_a$ and $R_b$, and satisfies the gauge transformation
\begin{equation}
\Psi_a=S\Psi_b  \ 
\end{equation}
in the overlapping region $R_{ab}$. 

The Laplacian operator in a curved space and in the presence of an external vector potential must be written in a covariant form replacing the spatial derivative $\partial_i$ by extended derivative $D_i=\partial_i-ieA_i$. Taking into account the electrostatic self-interaction the Schr\"odinger equation becomes:
\begin{equation}
\label{S}
\left[-\frac{\alpha^2}{2Mr^2}\partial_r(r^2\partial_r)+\frac{(\vec{L}_q^2
-q^2)}{2Mr^2}+\frac Kr\right]\Psi(\vec r)=E\Psi(\vec r) \ ,
\end{equation}
where $M$ is the mass of the particle and the conserved angular momentum operator is given by
\begin{equation}
\vec L_q=\vec r\times(\vec p-e\vec A)-q\hat r \ .
\end{equation}

In order to analyse the Schr\"{o}dinger equation above we shall adopt the
usual approach: We write the solution in the form
\begin{equation}
\label{Psi0}
\Psi(\vec r)=R_{q,l}(r)Y_{l,m}^q(\theta,\phi) \ ,
\end{equation} 
where $Y_{l,m}^q$ are the monopole harmonics\cite{Wu-Yang1}, solutions of the eigenvalues equation  
\begin{equation}
\vec L_q^2Y_{l,m}^q=l(l+1)Y_{l,m}^q \ ; \ L_zY_{l,m}^q=mY_{l,m}^q \ ,
\end{equation}
with $l=|q|, |q|+1, |q|+2, ...$ and $m=-l, -l+1, ..., l$. Substituting (\ref{Psi0}) into (\ref{S}) we obtain an ordinary differential equation to the radial function:
\begin{equation}
\frac1{r^2}\frac d{dr}\left(r^2\frac{dR_{q,l}}{dr}\right)-\frac{l(l+1)-q^2}
{\alpha^2r^2}R_{q,l}-\frac{2M}{\alpha^2}\left(\frac Kr-E\right)R_{q,l}=0 \ .
\end{equation} 

For $\alpha < 1$ the system presents only scattering states; however for $\alpha > 1$ the parameter $K$ becomes negative and bound states charged particle-monopole can be formed. Defining an effective orbital angular quantum number $\lambda_l^q=-1/2+\alpha^{-1}\sqrt{(l+1/2)^2-(q^2+a^2)}$ with $a^2=1/4(1-\alpha^2)$, the radial solutions of the above equation can be expressed in terms of hypergeometric functions as shown below.\\
$i)$ Bound states:
\begin{equation}
\label{R1}
R_{l,q}(r)=C_{l,q}r^{\lambda_l^q}{e^{-\kappa r}}_1F_1(-\gamma+\lambda_l^q+1,
2(\lambda_l^q+1);2\kappa r) \ ,
\end{equation}
with
$\kappa = \sqrt{-2ME/\alpha^2} \ \mbox{ and }  \gamma=-K \sqrt{-M/2\alpha^2E}$.

In order to have bound states we shall admit $E<0$, and the parameter $-\gamma+\lambda_l^q+1$ must obey the restriction below:
\begin{equation}
\gamma-\lambda_l^q-1=n \ ,
\end{equation}
$n$ being a non negative integer number. With the above condition the energy becomes discrete,
\begin{equation}
E_{n,l}^q=-\frac{MK^2}{2\alpha^2(\lambda_l^q+n+1)^2} \ .
\end{equation}

Comparing the spectrum above with similar one obtained in the absence of
magnetic charge\cite{Mello}, we can observe:\\
$a)$ When $q$ takes half-integer values, the ordinary orbital angular quantum numbers $l$ also assume half-integer values leading to a new series of the energy spectrum; however for integer values we can develop this comparison. Considering $q=1$ the relation between the energies of respective ground states is:
\begin{equation}
E_{0,1}^1/E_{0,0}=\left[1/2+\alpha^{-1}\sqrt{1+\alpha^2/4} \right]^{-2} \ ,
\end{equation}
which is smaller than unity for $\alpha > 1$. With this result we can infer that the system becomes less stable in the presence of a magnetic charge.\\
$b)$ The first excited state for $q=0$, happens for $n=0$ and $l=1$; however for $q=1$ the first exited state happens for $n=1$ up to $\alpha<\frac{2}{\sqrt{3}}$. For $\alpha>\frac{2}{\sqrt{3}}$ there is an inversion of the lines and the first exited state arises for $n=0$ and $l=2$.\\
\noindent\\
$ii)$ Scattering states:\\
\begin{equation}
\label{R2}
R_{l,q}(r)=C_{l,q}r^{\lambda_l^q}{e^{ikr}}_1F_1(\lambda_l^q+1+i\eta,
2(\lambda_l^q+1);-2ikr) \ ,
\end{equation}
with $k=\sqrt{2ME/\alpha^2} \mbox{ and } \eta=K\sqrt{M/2E\alpha^2}$.

As it is well known some of the most important informations about the interaction between a test particle and a target, are given by the scattering amplitude, $f(\theta)$, which by its turn depends on the phase shift, $\delta_l$.

\section{Scattering Amplitude}

Here we shall present the scattering amplitude for a charged particle interacting with an Abelian magnetic monopole in a global monopole background. So the scattering amplitude will contain informations about the gravitational and electromagnetic interactions.

The asymptotic behavior of the complete wavefunction associated with the scattering states can be obtained from (\ref{R2}):
\begin{equation}
\label{Psi}
\Psi(\vec{r})\sim 4\pi\sum_{l,m}a_{l,m}Y_{l,m}^q(\theta,\phi)
\frac{\sin[s(r)-\lambda_l^q\pi/2+\gamma_l]}{kr} \ ,
\end{equation}
where $s(r)=kr-\eta\ln(kr)$ and $\gamma_l=arg \Gamma(\lambda_l^q+1+i\eta)$ is the Coulomb phase shift with non-integer $\lambda_l^q$.

The constant $a_{l,m}$ can be determined considering (\ref{Psi}) representing an incoming distorted plane wave, $e^{i[\vec{k}\cdot\vec{r}+\eta\ln(kr-\vec{k}\cdot\vec{r})]}$, propagating in the $\Omega'$ direction, together with an outgoing scattering spherical wave. By using the completeness relation for the monopole harmonics\cite{Wu-Yang1}, we find:
\begin{equation}
\label{a}
a_{l,m}=e^{-i(\lambda_l^q\pi/2-\gamma_l)}(Y^q_{l.m}(-\Omega'))^* \ ,
\end{equation}
with $-\Omega'$ representing the angular variables $(\pi-\theta',\pi+\phi')$. Substituting (\ref{a}) into (\ref{Psi}) we get: 
\begin{eqnarray}
\label{Psi1}
\Psi(\vec{r})&\sim&\frac2{ik}\left\{\frac{e^{is(r)}}r\sum_{l=|q|}^\infty
e^{-i(\lambda_l^q\pi-2\gamma_l)}\sum_{m=-l}^{l}\left(Y_{l,m}^q(-\Omega')
\right)^*Y_{l,m}^q(\Omega)\right.\nonumber\\
&-&\left.\frac{e^{-is(r)}}r\delta(\Omega+\Omega')\right\} \ .
\end{eqnarray}

Wu and Yang\cite{Wu-Yang2} have derived some properties of monopole harmonics, including the generalization of spherical harmonics addition theorem. Following their procedure, the outgoing solution becomes
\begin{equation}
\Psi_{out}(\vec{r})\sim \frac{e^{is(r)}}re^{iq(\phi+\phi'+\pi)}f(\gamma) \ ,
\end{equation}
where we can identify the scattering amplitude as
\begin{equation}
\label{f}
f(\gamma)=\frac{\sqrt{\pi}}{ik}e^{-iq(R+R'-\pi)}\sum_{l=|q|}^\infty
\sqrt{2l+1}e^{-i(\lambda_l^q\pi-2\gamma_l)}Y_{l,-q}^q(\gamma,0) \ .
\end{equation}

It is important for us, as we shall see later, to express the monopole harmonics in terms of Jacobi polynomials\cite{Wu-Yang1}:
\begin{equation}
\label{Y1}
Y_{l,-q}^q(\gamma,0)=\frac1{2^q}\sqrt{\frac{2l+1}{4\pi}}(1+\cos\gamma)^q
P_n^{0,2q}(\cos\gamma) \ ,
\end{equation}
where $n=l-q$. For simplicity only let us take $\theta'=0$, consequently $\gamma=\pi-\theta$. Adopting this particular choice, we obtain
\begin{equation}
\label{F}
f(\theta)=\frac1{2ik}\left(\frac{1-x}2\right)^q\sum_{l=|q|}^\infty(2l+1)
e^{2i\delta_l}P_n^{2q,0}(x) \ ,
\end{equation}
with $x=\cos\theta$ and $\delta_l=(l-\lambda_l^q)\pi/2+\gamma_l$.

Although (\ref{F}) provides the complete scattering amplitude, we are unable to proceed the summation and obtain an analytical result. However, for small scattering angle, i.e., $\theta\ll 1$, the main contribution comes from large value of $l$. In this case it is possible to obtain an approximate closed expressions to (\ref{F}).

The phase shift in (\ref{F}) has two different contributions. The first one comes from the modification on the orbital angular quantum number due to the geometry of the spacetime and the magnetic interaction: $\delta_l^{(1)}=(l-\lambda_l^q)\frac{\pi}{2}$. The second one is the Coulomb phase shift $\delta_l^{(2)}=\gamma_l=arg\Gamma(\lambda_l^q+1+i\eta)$. Defining $z_l=l+\frac{1}{2}$, we can develop an expansion for both contributions in powers of $\frac{1}{z_l}$ and keep terms up to the first order only. The approximate expressions for the first contribution is
\begin{equation}
\label{delta1}
\delta_l^{(1)}\approx \frac\pi2\left[z_l(1-\alpha^{-1})+\frac{a^2+q^2}
{2\alpha z_l}+O\left(\frac1{z_l^2}\right)\right] \ .
\end{equation}
As to the Coulomb phase shift, after some steps we get:
\begin{equation}
\label{delta2}
\delta_l^{(2)}\approx\eta\left[\ln\left(\frac{z_l}\alpha\right)+\frac1{24z_l^2}
-\frac{q^2}{2z_l^2}+O\left(\frac1{z_l^4}\right)\right]+O(\eta^3) \ .
\end{equation}

So the total phase shift is given by the sum of (\ref{delta1}) with (\ref{delta2}). This result is:
\begin{equation}
\delta_l\approx\frac\pi2\left[z_l(1-\alpha^{-1})+\frac{a^2+q^2}{2\alpha z_l}\right] +\eta\left[\ln\left(\frac{z_l}\alpha\right)+O\left(\frac1{z_l^2}\right) \right] \ .
\end{equation}

Substituting the above phase shift in (\ref{F}), we get:
\begin{equation}
f(\theta)\approx\alpha^{-2i\eta}\left[f^{(0)}(\theta)+f^{(1)}(\theta)+
f^{(C)}(\theta)\right] \ ,
\end{equation}
where
\begin{equation}
f^{(0)}(\theta)=\frac1{ik}\left(\frac{1-x}2\right)^q\sum_{l=|q|}^\infty
z_le^{i\omega z_l}P_n^{2q,0}(x) \ ,
\end{equation}
\begin{equation}
f^{(1)}(\theta)=\frac{\pi(a^2+q^2)}{2\alpha k}\left(\frac{1-x}2\right)^q
\sum_{l=|q|}^\infty e^{i\omega z_l}P_n^{2q,0}(x) 
\end{equation}
and
\begin{equation}
f^{(C)}(\theta)=\frac{2\eta}k\left(\frac{1-x}2\right)^q\sum_{l=|q|}^\infty
z_l\ln(z_l)e^{i\omega z_l}P_n^{2q,0}(x) \ ,
\end{equation}
with $x=\cos\theta$ and $n=l-q$. 

Considering $q>1$, the two first contributions, $f^{(0)}$ and $f^{(1)}$, can be obtained using the generating function for the Jacobi polynomials. The results are:
\begin{equation}
\label{f0}
f^{(0)}(\theta)=-\frac{(1-\cos\theta)^q}{2\sqrt2 k}\frac{g_\omega(\theta)}
{(\cos\omega-\cos\theta)^{3/2}[(\cos\omega-\cos\theta)^{1/2}-i\sqrt2
\sin(\omega/2)]^{2q}} \ ,
\end{equation}
where
\begin{equation}
g_\omega(\theta)=\sin\omega+2q(\cos\omega-\cos\theta)^{1/2}\frac{[\sin\omega+
\sqrt2\cos(\omega/2)(\cos\omega-\cos\theta)^{1/2}]}{(\cos\omega-
\cos\theta)^{1/2}-i\sqrt2\sin(\omega/2)} \ ,
\end{equation}
and
\begin{equation}
\label{f1}
f^{(1)}(\theta)=\frac{\pi(a^2+q^2)}{2\sqrt2\alpha k}\frac{(1-\cos\theta)^q}
{(\cos\omega-\cos\theta)^{1/2}[(\cos\omega-\cos\theta)^{1/2}-i\sqrt2
\sin(\omega/2)]^{2q}} \ ,
\end{equation}
where $\omega=\pi(1-\alpha^{-1})$.

Following an analogous procedure as in Ref. \cite{Mello}, the Coulomb contribution can be expressed in a integral form:
\begin{equation}
\label{fc}
f^{(C)}(\theta)= -\frac{i\eta}{\sqrt2 k}\frac{g_\omega(\theta)}{(\cos\omega- \cos\theta)^{3/2}}\frac{(1-\cos\theta)^q\Theta_\omega(\theta)}{[(\cos\omega-\cos\theta)^{1/2}
-i\sqrt2\sin(\omega/2)]^{2q}} \ ,
\end{equation}
where
\begin{equation}
\Theta_\omega(\theta)=\int_0^\infty dx f_\omega(x,\theta) \ ,
\end{equation}
with
\begin{eqnarray}
\label{fw}
f_\omega(x,\theta)&=&\frac1{x^2}\left\{\sin(x)-\frac{(\cos\omega-\cos\theta)}
{g_\omega(\theta)}\left[\left(\frac{\cos\omega-\cos\theta}{\cos(\omega+x)-
\cos\theta}\right)^{1/2}\times\right.\right.\nonumber\\
&&\left(\frac{(\cos\omega-\cos\theta)^{1/2}-i\sqrt2
\sin(\omega/2)}{[\cos(\omega+x)-\cos\theta]^{1/2}-i\sqrt2\sin((\omega+x)/2)}
\right)^{2q}\nonumber\\
&-&\left(\frac{\cos\omega-\cos\theta}{\cos(\omega-x)-\cos\theta}\right)^{1/2}
\times\nonumber\\
&&\left.\left.\left(\frac{(\cos\omega-\cos\theta)^{1/2}-i\sqrt2
\sin(\omega/2)}{[\cos(\omega-x)-\cos\theta]^{1/2}-i\sqrt2\sin((\omega-x)/2)}
\right)^{2q}\right]\right\} \ .
\end{eqnarray}

Although the expression obtained for the total scattering amplitude is a long one, some relevant aspects of the scattering process can be observed: All its components, (\ref{f0}) - (\ref{fw}), present a singular behavior for $\theta$ close to $|\omega|=\pi|1-\alpha^{-1}|$. This especific behavior characterizes the physical signature of the global monopole. In principle, by measuring the scattering angle $\theta_0$ of the ringlike region where $f(\theta)$ becomes very large, it is possible to determine the parameter $\alpha$.

Some other interesting properties associated with the scattering amplitude can be observed: $(i)$ The first one is related with the purely magnetic scattering. Taking in our result $\alpha=1$, the most relevant component, $f^{(0)}(\theta)$, will be given by
\begin{equation}
f^{(0)}(\theta)=-\frac q{2k}\frac1{\sin^2(\theta/2)} \ ,
\end{equation}
consequently, the resulting differential cross section is 
\begin{equation}
\frac{d\sigma^{(0)}}{d\Omega}=\left(\frac q{2k}\right)^2
\frac1{\sin^4(\theta/2)}\ ,
\end{equation}
which coincides with the Rutherford formula for the Coulomb scattering for small angles. $(ii)$ For $\alpha<1$, $\omega=\pi(1-\alpha^{-1})<0$. In this case
\begin{equation} 
f^{(0)}(\theta)\to 0 \  for \ \theta\to 0 \ ,
\end{equation}
on the other hand for $\alpha>1$, $\omega $ becomes positive. So we have
\begin{equation}
f^{(0)}(\theta)\to\infty \ for \ \theta\to 0 \ .
\end{equation}
These two different behaviors are consequence of the factor $(1-\cos\theta)^q$ in the numerator of (\ref{f0}) to (\ref{fc}) and the change in the sign of the function $\sin\omega$ in their denominator.

\section{Concluding Remarks}

In this paper some of the most important results obtained are related with the analysis of the scattering amplitude associated with the charged particle-magnetic monopole in the global monopole background. As we have shown, this function presents an universal singular behavior at the ringlike region $\theta=\pi|1-\alpha^{-1}|$; moreover depending if the parameter $\alpha$ assumes values smaller or bigger than unity, this function presents a
completely different behavior at $\theta=0$. Finally we can say that for the case when $\alpha>1$, bound states of charged particle-magnetic monopole system can be formed, due to the attractive electrostatic self-interaction with the energy spectrum depending on the value of the parameter $q=eg$.
\\
\\
{\bf{Acknowledgments}}\\
We would like to thank Conselho Nacional de Desenvolvimento Cient\'\i fico e 
Tecnol\'ogico (CNPq.) and CAPES for partial financial support.

\end{document}